\documentstyle[11pt,psfig,aaspp4,flushrt]{article}
\begin{document}


\title{The Butcher-Oemler Effect at Moderate Redshift}

\author{Anne J. Metevier\altaffilmark{1}}
\affil{Department of Physics and Astronomy, Dearborn Observatory, Northwestern University, 2131 Sheridan Road, Evanston, IL 60208-2900\\
metevier@lilith.astro.nwu.edu}

\altaffiltext{1}{present address:\\
Department of Astronomy and Astrophysics, University of California, Santa Cruz, CA 95064\\
anne@ucolick.org}

\altaffiltext{2}{Visiting Astronomer, Kitt Peak National Observatory, National Optical Astronomy Observatories, which is operated by the Association of Universities for Research in Astronomy (AURA), Inc., under cooperative agreement with the National Science Foundation.}

\author{A. Kathy Romer\altaffilmark{2}}
\affil{Department of Physics, Carnegie Mellon University, 5000 Forbes Ave., Pittsburgh, PA 15213-3890\\
romer@astro.phys.cmu.edu}


\author{M. P. Ulmer\altaffilmark{2}}
\affil{Department of Physics and Astronomy, Dearborn Observatory, Northwestern University, 2131 Sheridan Road, Evanston, IL 60208-2900\\
m-ulmer2@nwu.edu}

\begin{abstract}
We present the results of Butcher-Oemler-style analysis of three moderate-redshift ($0.1<z<0.2$) clusters which have bimodal X-ray surface brightness profiles.  We find that at least two of these clusters exhibit unusually high fractions of blue galaxies ($f_{b}$s) as compared to clusters at comparable redshifts studied by Butcher and Oemler (1984).  This implies that star formation is occurring in a high fraction of the galaxies in the two clusters.  Our results are consistent with hierarchical clustering models in which subcluster-subcluster mergers create shocks in the intracluster medium.  The shocks, in turn, induce simultaneous starbursts in a large fraction of cluster galaxies.  Our study therefore lends weight to the hypothesis that the Butcher-Oemler effect is an environmental, as well as evolutionary, phenomenon.
\end{abstract}




\section{Introduction}

Studies of galaxy clusters are essential to the field of observational cosmology because they allow us to probe the evolution of structure in the universe.  One of the first extensive studies of cluster evolution was made by Butcher and Oemler (1984; hereafter BO84), who compared the colors of galaxies in 33 clusters over the redshift range $0 \leq z \leq 0.54$.  They calculated the fraction of blue galaxies in each cluster, defining ``blue'' galaxies as those whose B--V colors lay at least 0.2 magnitudes below the cluster's E/S0 ridgeline.  In so doing, Butcher and Oemler determined that the fraction of blue galaxies ($f_{b}$) in clusters increases with increasing redshift.  Specifically, they found that below $z=0.1$, cluster blue fractions are nearly negligible ($f_{b} \sim 0.03 \pm 0.01$); however, at $z>0.1$, cluster blue fractions increase steadily until $f_{b} \sim 0.25$ at $z=0.5$.  This trend has become generally known as the Butcher-Oemler effect ({\it cf.} Figure 3 of BO84; also see our results compared to BO84's in Section 5 below).  

\subsection{Establishing a Trend}

Many spectroscopic studies have been carried out to further investigate the Butcher-Oemler effect.  Dressler and Gunn (1982, 1983) found that the spectra of confirmed blue cluster members exhibited features similar to ``active" galaxies.  Lavery and Henry (1986) determined that the spectra of blue galaxies in BO84 clusters are dominated by features due to young- or intermediate-aged stars.  They suggested that the ``blueness" of the cluster galaxies is therefore caused by star formation, as it is known that young, hot O and B stars are brightest in the UV.  Couch and Sharples (1987) also provided spectral evidence that blue galaxies in $z \sim 0.3$ clusters have experienced a burst of star formation within 0.1 and 1.5 Gyr of the observed epoch.  Further spectroscopic studies that have confirmed the Butcher-Oemler effect and provided evidence that blue cluster members are actively forming stars include 
Dressler, Gunn, and Schneider (1985); Henry and Lavery (1987); Mellier et al. (1988); and Fabricant, McClintock, and Bautz (1991).  

Morphological studies of blue cluster members have strengthened these findings.  Thompson (1988) imaged four blue members of BO84 clusters and found that they were all disk systems.  Couch et al.'s (1994) combined spectroscopic and morphological study of two BO84 clusters at $z \sim 0.3$ also showed that the star-forming cluster members tend to be disk galaxies.  Lavery and Henry (1994), in a study of 19 BO84 cluster galaxies, determined that 50\% of the blue cluster members are disk dominated or have disk components.  More recent morphological investigations done with HST images confirm that star-forming members of high-redshift clusters tend to be spirals, many of which are undergoing interactions with other cluster galaxies.  Such studies include Dressler et al. (1994); Wirth, Koo, and Kron (1994); Wirth (1997); Andreon, Davoust, and Heim (1997); Lubin et al. (1998); van Dokkum et al. (1998); and Caldwell, Rose, and Dendy (1999).  In combination, these studies suggest that the evolutionary color trend observed by BO84 is indicative of (a) evolution of the stellar populations of cluster galaxies and/or (b) evolution of the morphologies of cluster galaxies.  

It should be noted, however, that the BO84 cluster sample was not selected in an optimum way.  Kron (1993) has pointed out that BO84's sole selection criterion was to require spectroscopically determined cluster redshifts.  At that time (1984), large, objectively-selected cluster samples with complete redshift follow-ups, such as the EDCC (Lumsden et al. 1992) and the EMSS (Gioia et al. 1990), were not available.   Therefore, Butcher and Oemler (1984) were forced to use cluster redshifts from a variety of sources.  As a consequence, their sample was heterogeneous and not fully representative of the cluster population as a whole.  Simply put, it is possible that many of the clusters studied in BO84 are extraordinary rather than representative.
 
To counteract biasing due to selection effects, Annis (1994) undertook a study of two samples of X-ray-selected EMSS clusters, one at $z=0$ and another at $z=0.33$.  Although his results did not imply an evolutionary trend as striking as BO84's, Annis determined that the $z=0.33$ sample of clusters exhibits a wide range of blue fractions $0<f_{b}<0.5$.  According to Annis, the measured range of blue fractions indicates that many galaxies in a cluster may undergo simultaneous starbursts, possibly due to the merging of a subcluster with the main cluster.  More recently, Lubin (1996) ran 57 rich Palomar Distant Cluster Survey clusters (covering a redshift range $0.2 \leq z \leq 1.2$) through a Butcher-Oemler-style analysis.  While there was significant scatter in the results, Lubin concluded that cluster blue fractions do increase with increasing redshift.  Lubin also studied clusters at much higher redshifts than BO84 and found that at $z>0.6$, cluster blue fractions are as high as 0.4.  

\subsection{Discerning the Effect} 

The goal of this paper is to investigate the cause of the BO effect.  Several studies have suggested possible interpretations of the BO84 findings, beginning with a study done by Butcher and Oemler themselves. In an early investigation of the blue fractions of two distant clusters, Butcher and Oemler (1978) noted that there was a similarity between the number of blue galaxies in high-$z$ clusters and the number of S0 galaxies in nearby clusters.  They therefore suggested that blue galaxies in young clusters are rich in gas and young stars, but evolve to form S0s such as those in present-day clusters.  Very little star formation would occur in the evolved S0's; their gas would have been tidally stripped or exhausted by earlier star formation.  

Another potential explanation for the BO trend is that the blue cluster members BO84 studied are field galaxies falling into the cluster for the first time.  Ram pressure from the intracluster medium (ICM) would induce star formation in these infalling galaxies (Evrard 1991).  Observational evidence to support this theory comes from studies which have shown that blue galaxies predominantly lie in the outer edges of clusters (Rakos, Odell, \& Schombert 1997), while ellipticals decline in prevalence in these outer, less dense, regions (Dressler 1980; Smail et al. 1997).  A third possibility is that the blue cluster galaxies are involved in galaxy-galaxy interactions.  Studies have shown that mergers may be responsible for the Butcher-Oemler effect in clusters at redshifts up to $z=0.4$ (Lavery \& Henry 1988; Lavery, Pierce, \& McClure 1992).  However, the porbability of galaxy-galaxy mergers is low in clusters due to cluster galaxies' high velocity dispersions.  Moore et al. (1996) suggest galaxy ``harassment", in which cluster galaxies ``brush against" each other many times at high speed, as a more likely evolutionary effect.  Moore et al. have shown through high-resolution numerical simulations that such high-speed encounters create starbursts in cluster spirals and pull material away from the spirals until very little gas is left and the spirals are transformed into a dwarf spheroidals.

Recently, it also has been theorized that the Butcher-Oemler effect may be a reflection of the fact that clusters formed via a hierarchical model (e.g. Serna \& Gerbal 1996) in which smaller units---individual galaxies, then groups and subclusters---were attracted and merged into the main cluster under the influence of gravity.  Within the context of this model, Kauffmann (1995) suggested an explanation for the BO trend: subclustering would have been more common in the past, and the merging of subclusters would produce shocks in the ICM, which would in turn induce simultaneous star formation in a large fraction of galaxies in a cluster.  According to this interpretation, galaxy clusters with high degrees of subclustering would tend to have high fractions of blue galaxies irrespective of their redshifts.  A recent theoretical study provides support for this theory: Bekki (1999) has performed an $n$-body simulation detailing the evolution of a spiral galaxy within a group that merges with a cluster.  Bekki has found that due to the merger, gas within the spiral is transferred to its central region and undergoes a starburst.

Observational support comes from Caldwell and Rose (1997), who gathered spectra of galaxies in five nearby ($0.01<z<0.06$) clusters that are known to have varying degrees of substructure.  Their data provide evidence that starbursts are induced in subcluster galaxies as they pass through the center of a main cluster, enhancing the cluster's blue fraction.  These results agree with the explanations presented in Kauffmann (1995) and Bekki (1999) of the physical processes behind the BO effect.  

Here we present the results of our own test of the influence of hierarchical clustering on cluster blue galaxy fractions.  Our observations were based on the hypothesis that clusters with substructure should exhibit high blue fractions, regardless of their redshifts.  Our study differs from that of Caldwell and Rose (1997) in that our clusters were chosen because of their X-ray morphologies, which suggest that the clusters are dynamically young.  In Section 2, we discuss our selection of three nearby clusters ($0.1<z<0.2$) which exhibit noticeable substructure in their X-ray emission.  We detail our observations of the clusters in Section 3, and in Section 4 we describe our reduction and analysis techniques, which were designed to closely follow those used by BO84.  Lastly, in Section 5, we present the blue fractions of our three clusters and discuss their significance. 

Throughout this paper, we use values of $H_{0}=50$ $km$ $s^{-1}$ $Mpc^{-1}$ and $q_{0}=0.1$ as was done in BO84.

\section{Our Cluster Sample}

We have chosen to study Abell 98, Abell 115, and Abell 2356 because each of these clusters clearly exhibits substructure.  Forman et al. (1981) first studied A98 and A115 at X-ray wavelengths with the Einstein IPC; in the case of A98 they found two distinct X-ray emitting regions separated by nearly 9 arcminutes---A98N and A98S (see Figure 1).  Beers (1983) later found that this cluster's galaxy distribution closely follows its X-ray surface brightness distribution.  Dynamical studies of A98N and A98S have shown that there is a 98\% probability that the two regions are bound and moving toward one another with a relative velocity of 455 $km$ $s^{-1}$ (Beers, Geller, \& Huchra 1982; Krempec-Krygier \& Krygier 1995).  The southern region also contains an extended radio source; the associated jets are thought to be influenced by the intracluster medium (ICM) which itself may be disturbed by a subcluster-subcluster merger (Krempec-Krygier \& Krygier 1995). 

Abell 115 also has a bimodal X-ray surface brightness distribution (Forman et al. 1981; Pierre \& Starck 1998; see Figure 2).  Beers, Geller, and Huchra (1983) found that this cluster's optical galaxy distribution contains three significant clumps, two of which are associated with the bright X-ray regions.   A115 also contains a radio source with distorted jets (Gregorini \& Bondi 1989).  Abell 2356 was observed by Ulmer and Cruddace (1982) using Einstein; they found that A2356 is ``boomerang"-shaped in its X-ray structure (see Figure 3).

Besides their X-ray substructure, A98, A115, and A2356 have been chosen because of their low redshifts ($0.1<z<0.2$).  This makes them ideal for our study: if our hypothesis is correct, these clusters will exhibit much higher blue fractions than those studied in BO84 at comparable redshifts.  There are several observational advantages to using low-redshift clusters for this study.  They are fully resolved in the X-ray by both Einstein and ROSAT; it is also easier to detect intrinsically fainter cluster galaxies in optical CCD images, foreground/background galaxy contamination is less of a problem, and $k$-corrections are less severe.  Properties of the clusters in our sample, including their Abell richness classes and X-ray luminosities, are listed in Table 1.

\section{Observations and Data Reduction}

We observed A98, A115, and A2356 in B, V, and R using the CCD Imager camera on the 0.9-meter telescope at Kitt Peak National Observatory.  This camera was optimal for our study because of its large field of view: all of our images are 23.21 arcminutes to a side.  Images were taken over two nights on August 31 and September 1, 1995.  We also imaged reference fields at an offset of 18 arcminutes from the clusters; we used the reference images to correct cluster observations for line-of-sight (``background'') galaxy contamination.   We imaged a fourth, spherically symmetric ``control'' cluster (Abell 2390), but due to poor weather conditions these control data were not of photometric quality and could not be included in our study.

To begin reduction, we trimmed, bias subtracted and flat-field corrected each image using the {\tt ccdproc} task in the IRAF software package.  Dome flats were used in order to remove pixel-to-pixel variations introduced by the CCD in our images; smoothed twilight sky flats were used to remove large-scale gradients. We then performed object detection, classification, and photometry on the images using the FOCAS software package (Valdes 1989).  

Following the methodology of Bernstein et al. 1995 (hereafter B95), we made an investigation into the star/galaxy separation done by FOCAS.  The technique makes use of two different magnitudes FOCAS calculates for each object: $m_{core}$ and $m_{tot}$.  The core magnitude ($m_{core}$) of a detected object is determined by calculating the flux of its nine brightest central pixels.  FOCAS determines an object's total magnitude ($m_{tot}$) by expanding the object's boundaries several pixels in all directions, then measuring the flux above the sky level in this larger region.  B95 proved through Monte Carlo simulations that $m_{tot}$ is a good estimate of an object's true magnitude.  The classification test developed by B95 then involves comparing an object's total magnitude ($m_{tot}$) with its extent ($m_{core}-m_{tot}$).  In this test, one would expect a star, represented by a well-defined point spread function, to have an extent of nearly zero.  In contrast, a galaxy should have a higher extent.  Figure 4 shows that stars and galaxies are clearly distinguished in our V band images.  Vertical dashed lines in the plots denote the apparent magnitudes where $M_{V}=-20$; our BO analysis includes only galaxies with $M_{V}<-20$ (see section 4).

We calibrated our photometry by comparing FOCAS total magnitudes to true magnitudes for standard stars at varying airmasses, taking exposure time into account.  We determined extinction coefficients (to be used in airmass corrections) and zeropoints simultaneously by plotting airmass versus $m_{inst} - m_{true}$ for our standards, where $m_{inst}$ is an object's intrumental magnitude.  Then we fit a straight line to our data points; the slope of this line represents the extinction coefficient, and the zeropoint is represented by the y-intercept.  Our $1\sigma$ photometric errors were less than 0.05 magnitudes.

We then corrected for galactic extinction.  As was done in BO84, we used the mean of extinction values calculated from the relations provided in Sandage (1973) and de Vaucouleurs, de Vaucouleurs, and Corwin (1976).  We also $k$-corrected our data using the same method as BO84, by interpolating $k$-correction values from the values given for E/S0 galaxies in Pence (1976).  Galactic extinction and $k$-correction values are given in Tables 2 and 3, respectively, for each cluster.

\section{Analysis}

We made every effort to analyze our data in a manner as similar as possible to the BO84 analysis.  In order to create lists of cluster galaxies, we first fit an ellipse to each cluster to define the portion of sky it covered in our images.  To fit our cluster ellipses, we used a variation on the method used by Annis (1994), who studied the Butcher-Oemler effect in spherically symmetric X-ray clusters.  Annis fit circles to his clusters, defining the center of each cluster as the position of the brightest galaxy in that cluster.  This definition was based on the finding by Beers and Tonry (1986) that massive, central D galaxies mark cluster centers nearly as well as the clusters' X-ray centroids.  Annis set the radii of his clusters at 0.5 Mpc, following the suggestion of Hill and Lilly (1991).

Because our clusters are bimodal rather than spherically symmetric, we modified Annis's circle-fitting method to an ellipse-fitting technique.  For a given cluster, we determined two focii rather than a single cluster center.  These focii were defined to coincide with the positions of the brightest galaxies in each of the cluster's two constituent subclusters.  Secondly, we defined the semimajor axis of each cluster ellipse to be half the distance between the two focii plus 0.5 Mpc, where the addition of 0.5 Mpc was done in order to be consistent with the radius of Annis's circle-fitting method.  Galaxies which fell within the ellipses we defined were considered cluster galaxies; galaxies outside of the ellipses were not included in our lists.  We also ran background galaxy subtraction---see the description below. 

Once we created cluster galaxy lists, we corrected all of the galaxies in the lists (including late-types) for the color-magnitude effect.  BO84 also did this.  The color-magnitude effect is a linear relation for early-type galaxies: more luminous early-type galaxies tend to have ``redder'' colors than less luminous early types.   This effect was particularly evident in the color-magnitude diagrams of A98 and A2356.  We determined the color-magnitude relation for these two clusters by $\chi^{2}$ fitting to the distribution of galaxies in the $V$ versus $B-V$ plane.  The slopes of the relations for both clusters were the same.  When doing their color-magnitude slope fitting, BO84 found that their slopes agreed with those presented in Visvanathan \& Sandage (1977).   Ours did not; however, we found that our blue fraction results were consistent within the errors whether we used our fit color-magnitude slope or the slope calculated by Visvanathan \& Sandage.  Therefore, like BO84, we used the latter for all three clusters.  We then made a magnitude cut to our data.  Following BO84, we only included galaxies in our analysis which were brighter than $M_{V}=-20$, as there is evidence that this cutoff corresponds to the magnitude limit between dwarf and normal galaxies (Pierce 1988).  

We measured the number of blue galaxies inside our cluster ellipses, using the same blue galaxy definition as BO84: a galaxy is ``blue'' if it is at least 0.2 magnitudes bluer than the ridgeline of E/S0 galaxies in the cluster.  In Figure 5, we show the background-subtracted, color-magnitude-corrected color distribution of galaxies in each of our clusters.  One can see that we have well sampled the E/S0 ridgelines of A98 and A2356, while the ridgeline is less apparent for A115.  Color-magnitude diagrams for each of our clusters are given in Figure 6.  Our blue galaxy cutoff is also plotted in Figure 6.

Our last analysis step was to remove contamination by foreground and background galaxies.  In order to do this, we took images of the field near each cluster (see section 3).\footnote{Due to time constraints, no field images were taken for Abell 115.  Instead, the field image from Abell 98 was used to make a correction for this cluster.}  We reduced field images in the same manner as the cluster images; field galaxies were corrected for galactic extinction and $k$-corrected as if they were at the redshift of the cluster.  As with the galaxies in the cluster images, we corrected for the color-magnitude effect, we made a $M_{V}<-20$ magnitude cut, and we again used the BO84 definition of a blue galaxy to determine the number of blue field galaxies.

We corrected the total number of galaxies $N$ in each cluster by counting the number of galaxies in the field image, multiplying by a normalization factor\footnote{normalization factor = (area of cluster ellipse)/(area of field image)}, then subtracting the resulting number from the total cluster count $N$.  An equivalent approach was used to correct the number of blue galaxies in each cluster $N_{b}$.  Finally, we calculated blue fractions for each cluster from the following equation (Annis 1994):
\begin{equation}
f_{b} = \frac{\rm N_{b}}{\rm N} \pm \sqrt{\frac{\rm (N_{b})\left(1-\frac{\rm N_{b}}{\rm N}\right)}{\rm N}}
\end{equation}

\noindent The error in this equation is based on counting statistics for a binomial distribution.  The blue fractions we determined are given in Table 4.  Also included in Table 4 are the results of our analysis of A98, treated as two separate clusters (A98N and A98S), and BO84's blue fraction results for two clusters with low redshifts and X-ray indications of substructure (A1367 and Hercules).  These results with be discussed in the next section.

\section{Results and Discussion}

Our blue fraction results are plotted against BO84's in Figure 7 and listed in Table 4.  We have shown that bimodal X-ray clusters tend to be bluer than expected from the ``guiding line" presented by BO84 against their data.  In particular, A98 and A115 exhibit impressive blue fractions.  When we extended our analysis of A98 and treated it as two separate, symmetric clusters, we found that a large percentage (60\%) of the galaxies in A98N are blue (see Table 4).  While it is true that we have small number statistics for this subcluster, it is possible that the majority of the galaxies in A98N are undergoing star formation.
  
Before going on to discuss the implications of our results, we would like to stress their validity.  First, we consider the fact that our blue fraction results may be affected by photometry errors.  Our airmass corrections are color neutral and errors here would not change the cluster and field galaxy colors we calculated.  The airmass correction does change our $M_{V}$ cutoff, but our results are insensitive to changes in $M_{V}$ within the uncertainties of the airmass correction.  Indeed, we have found by rerunning our analysis that our $1\sigma$ photometry error of 0.05 (which also includes zero-point uncertainties) gives rise to changes in our derived blue fractions that are well within the statistical error bars.  Second, we note that our $k$-corrections and galactic extinction corrections were small enough that they too changed our results by less than the statistical uncertainties.  Third, we have tested our results against errors in our color-magnitude correction.  We have varied the slope of this correction 1$\sigma$ high and 1$\sigma$ low, and we have found that the affect of this change on our blue fraction results in well within the counting errors.  

Finally, we ran two separate tests to ensure that our choice of cluster boundaries has not affected our blue fractions.  This is important, because if we included too many field galaxies in a cluster list, we would expect the cluster blue fraction to be very high: a higher percentage field galaxies have ``blue'' colors than do cluster galaxies.  We first tested the positions of the cluster ellipses we derived by moving each ellipse back and forth in the x- and y-directions by up to 15\% of the distance between its focii.  We then tested the sizes of the cluster ellipses by enlarging and reducing the sizes of their semi-major axes up to 15\%.  We chose these 15\% shifts because it was obvious, from looking at the data, that any shifts larger than this would have excluded, or included, too much data.  Again, as in the previous checks we have discussed, the revised blue fractions were well within statistical counting errors.  We conclude that statistical errors are the dominant source of uncertainty to our blue fractions and proceed on that basis.  

Our results give qualitative support to the hypothesis that hierarchical clustering is a physical process from which the Butcher-Oemler effect arises.  Indeed,  the BO84 data itself provides similar evidence: two of the BO84 clusters at low redshift exhibit very high blue fractions as well as X-ray substructure.  Abell 1367, at $z=0.0213$ and $f_{b}=0.19 \pm 0.03$, has two clear subcomponents in the X-ray (e.g. ASCA image from Churazov et al. 1996).  This cluster is most likely undergoing a cluster-subcluster merger:  in a recent study of X-ray data from both ASCA and ROSAT, Donnelly et al. 1998 showed that there is a strong shock in A1367's ICM.  Shocks are an expected biproduct of such mergers as shown by recent $n$-body simulations (e.g. Schindler \& M\"{u}ller 1993).  Hercules (Abell 2151), at $z=0.0371$, also exhibits a higher than average blue fraction in BO84: $f_{b}=0.14 \pm 0.02$.  This cluster shows evidence for substructure as well.  Bird, Davis, and Beers (1995) compared X-ray and optical data from Hercules.  They reported there is evidence that Hercules has at least three major components in its galaxy distribution.  They also detected weak X-ray emission from two of the three subclusters.   

The blue fractions BO84 derived for A1367 and the Hercules cluster (see Figure 7 and Table 4) are consistent with the results of our study.  Further evidence comes from Wang and Ulmer (1997), who have shown that there is a strong correlation between the X-ray ellipticities of ten of the BO clusters\footnote{Wang and Ulmer did not include A1367 or Hercules in their (1997) study.} and the blue fractions BO84 derived for them.  According to Wang and Ulmer, more elongated clusters tend to exhibit higher blue fractions.  However, the reader should also keep in mind that we do not have a complete investigation of the BO84 sample and so there may well be BO84 clusters that have substructure but do not have anomalously high blue fractions.  

We compare the X-ray luminosities of our clusters to those of fifteen BO84 clusters studied by Ebeling et al. (1998) in Figure 8.  From Figure 8(a), it is apparent that in general the cluster luminosities increase with redshift.  This is what one would expect for a flux-limited sample such as Ebeling et al. (1998).  Note that Abell 98 and Abell 2356 have X-ray luminosities comparable to that of Virgo ($L_{X}=1.00\times 10^{44}$ erg s$^{-1}$) whereas the X-ray luminosity of Abell 115 is more similar to that of Coma ($L_{X}=7.26\times 10^{44}$ erg s$^{-1}$).  In Figure 8(b), it can be seen that there is not an obvious correlation between cluster X-ray luminosities and blue fractions.

In light of the results of this study, it will be interesting to extend our work.  New software is now available to extract the structure of cluster galaxies, so that the colors of the disk and bulge components can be separated.  This way it will be possible for us to determine where star formation is occurring within the galaxies.  X-ray data will also play a prominent role in our future work.  There should be a strong correlation between cluster X-ray luminosities and their blue fractions: Kauffmann (1995) showed that in simulations the Butcher-Oemler effect was considerably more pronounced in more massive (more luminous) clusters than in less massive ones.  Moreover, spatially resolved temperature maps (from either AXAF or XMM) will highlight regions of shocked intracluster gas, and show the relationship between the metallicity of the ICM and the presence of blue cluster galaxies.  If a higher than average metallicity is found near the blue galaxies, it will provide a direct link between cluster galaxy evolution and the origin of the ICM.

In conclusion, we have determined that at least two of the three bimodal X-ray clusters we have studied exhibit unusually high blue fractions.  This work, coupled with that of Caldwell and Rose (1997), suggests that cluster galaxy evolution is a continuing process that does not stop at low redshift.  Further studies will strengthen our understanding of hierarchical clustering and its effects on cluster galaxy evolution.

\acknowledgments
We are grateful to Francisco Castander for advice on photometry and Frank Valdes for very patiently steering us around problems with FOCAS.  Jim Annis provided us with very useful discussions; his PhD thesis was a model for our study.  We are also indebted to Scott Severson for comments on earlier drafts of this paper.  This work was supported, in part, by funding from the NASA Space Grant to the State of Illinois.  This research has make use of the NASA/IPAC Extragalactic Database (NED) which is operated by the Jet Propulsion Laboratory, Caltech, under contract with the National Aeronautics and Space Administration.

\newpage

\references

Andreon, S., Davoust, E., \& Heim, T. 1997, A\&A, 323, 337

Annis, J.T. 1994, PhD Thesis, University of Hawaii

Arnaud, K.A. 1996, ADASS, 5, 17

Arnaud, M. \& Evrard, A.E. 1999, \mnras, 305, 631

Beers, T.C. 1983, PhD Thesis, Harvard University

Beers, T.C., Geller, M.J., \& Huchra, J.P. 1982, \apj, 257, 23

Beers, T.C., Geller, M.J., \& Huchra, J.P. 1983, \apj, 264, 356

Beers, T.C. \& Tonry, J.L. 1986, \apj, 300, 557

Bekki, K. 1999, \apj, 510, 15

Bernstein, G.M., Nichol, R.C., Tyson, J.A., Ulmer, M.P., \& Wittman, D. 1995, \aj, 110, 150

Bird, C.M., Davis, D.S., \& Beers, T.C. 1995, \aj, 109, 920

Butcher, H. \& Oemler, A. 1978, \apj, 219, 18

Butcher, H. \& Oemler, A. 1984, \apj, 285, 426

Caldwell, N. \& Rose, J.A. 1997, \aj, 113, 492

Caldwell, N., Rose, J.A., \& Dendy, K. 1999, \aj, 117, 140

Churazov, E., Gilfanov, M., Forman, W., \& Jones, C. 1996, Proc. ``R\"{o}ntgenstrahlung from the Universe'', eds. Zimmermann, H.U., Tr\"{u}mpler, J., and Yorke, H., MPE Report 263, 573

Couch, W.J. \& Sharples, R.M. 1987, \mnras, 229, 423

Couch, W.J., Ellis, R.S., Sharples, R.M., \& Smail, I. 1994, \apj, 430, 121

Donnelly, R.H., Markevitch, M., Forman, W., Jones, C., David, L.P., Churazov, E., \& Gilfanov, M. 1998, \apj, 500, 138

Dressler, A. 1980, \apj, 236, 351

Dressler, A. \& Gunn, J.E. 1982, \apj, 263, 533

Dressler, A. \& Gunn, J.E. 1983, \apj, 270, 7

Dressler, A., Gunn, J.E., \& Schneider, D.P. 1985, \apj, 294, 70

Dressler, A., Oemler, A., Butcher, H.R., \& Gunn, J.E. 1994, \apj, 430, 107 

Ebeling, H., Edge, A.C.. B\"{o}hringer, H., Allen, S.W., Crawford, C.S., Fabian, A.C., Voges, W., \& Huchra, J.P. 1998, \mnras, 301, 881

Evrard, A.E. 1991, \mnras, 248, 8

Fabricant, D.E., McClintock, J.G., \& Bautz, M.W. 1991, \apj 381, 33

Forman, W., Bechtold, J., Blair, W., Giacconi, R., Van Speybroeck, L., \& Jones, C. 1981, \apjl, 243, L133

Gioia, I.M., Maccacaro, T., Schild, R.E., Wolter, A., Stocke, J.T., Morris, S.L., \& Henry, J.P. 1990, \apjs, 72, 567

Gregorini, L. \& Bondi, M. 1989, A\&A, 225, 333

Henry, J.P, \& Lavery, R.J. 1987, \apj 323, 473

Hill, G.J. \& Lilly, S.J. 1991, \apj, 367, 1

Kauffmann, G. 1995, \mnras, 274, 153

Krempec-Krygier, J. \& Krygier, B. 1995, A\&A, 296, 359

Kron, R.G. 1993, in {\it The Deep Universe}, eds. B. Binggeli \& R. Buser (New York: Springer), p. 300

Lavery, R.J. \& Henry, J.P. 1986, \apj, 304, 5

Lavery, R.J. \& Henry, J.P. 1988, \apj, 330, 596

Lavery, R.J. \& Henry, J.P. 1994, \apj 426, 524

Lavery, R.J., Pierce, M.J., \& McClure, R.D. 1992, \aj, 104, 2067

Lubin, L.M. 1996, \aj, 112, 23

Lubin, L.M., Postman, M., Oke, B., Ratnakunga, K.U., Gunn, J.E., Hoessel, J.G., \& Schneider, D.P. 1998, \aj, 116, 584

Lumsden, S.L., Nichol, R.C., Collins, C.A., \& Guzzo, L. 1992, \mnras, 258, 1

Mellier, Y., Soucail, G., Fort, B., Mathez, G., \& Cailloux, M. 1988  A\&A 191, 19

Moore, B., Katz, N., Lake, G., Dressler, A., Oemler, A. 1996, Nature, 379, 613

Pence, W. 1976, \apj, 203, 39

Pierce, M.J. 1988, PhD Thesis, University of Hawaii

Pierre, M. \& Starck, J.-L. 1998, A\&A, 330, 801

Rakos, K.D., Odell, A.P., \& Schombert, J.M. 1997, \apj, 490, 194 

Sandage, A. 1973, \apj, 183, 711

Schindler, S. \& M\"{u}ller, E. 1993, A\&A, 272, 137

Serna, A., \& Gerbal, D. 1996, A\&A, 309, 65

Smail, I., Dressler, A., Couch, W.J., Ellis, R.S., Oemler, A., Butcher, H., \& Sharples, R.M. 1997, \apjs, 110, 213

Thompson, L.A. 1988, \apj, 324, 112

Ulmer, M.P. \& Cruddace, R.G. 1982, \apj, 258, 434

Valdes, F. 1989, in Proceedings of the 1st ESO/ST-ECF Data Analysis Workshop, edited by P.J. Grosbol, F. Murtagh, \& R.H. Warmels (ESO, Garching)

van Dokkum, P.G., Franx, M., Kelson, D.D., Illingworth, G.D., Fisher, D., \& Fabricant, D. 1998, \apj, 500, 714

de Vaucouleurs, G., deVaucouleurs, A. \& Corwin, H.G. 1976, {\it Second Reference Catalogue of Bright Galaxies} (Austin: University of Texas)

Visvanathan, N. \& Sandage, A. 1977, \apj, 216, 214

Wang, Q.D. \& Ulmer, M.P. 1997, \mnras, 292, 920

White, S.M., Briel, U.G., \& Henry, J.P. 1993, \mnras, 281, 8

Wirth, G.D. 1997, PASP, 109, 344

Wirth, G.D., Koo, D.C., \& Kron, R.G. 1994, \apj, 435, 105

\newpage

\figcaption[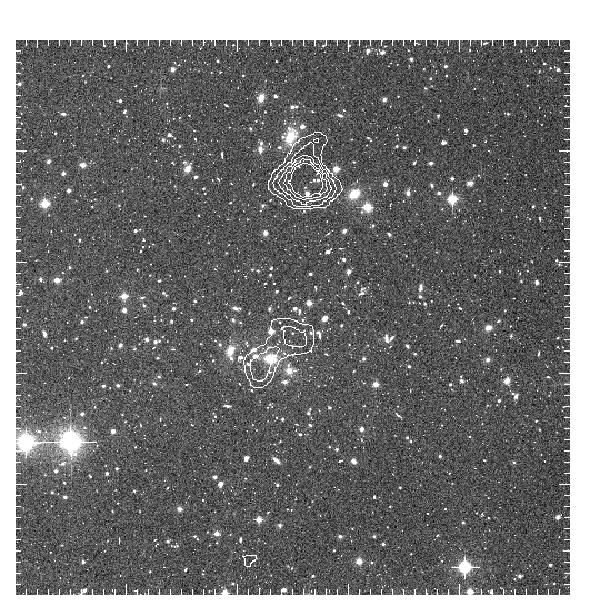]{Abell 98 R image with X-ray overlay.  Image dimensions are $23.21 \times 23.21$ arcminutes.  The optical image was taken with the 0.9-meter telescope at KPNO; the X-ray image is from Einstein archives.  Contour levels were chosen for consistency with Forman et al. 1981; levels are plotted at 5.2, 7.0, 8.8, 10.8, and 12.6 counts.}

\figcaption[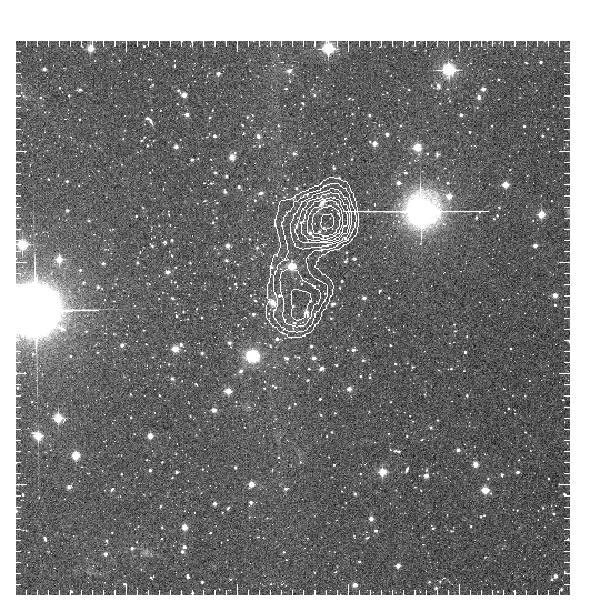]{Abell 115 R image with X-ray overlay.  Image dimensions are $23.21 \times 23.21$ arcminutes. The optical image was taken with the 0.9-meter telescope at KPNO; the X-ray image is from Einstein archives.  Contour levels were chosen for consistency with Forman et al. 1981; levels are plotted at 5.3, 7.7, 10.5, 13.3, 16.1, 18.9, 22.1, and 27.3 counts. }

\figcaption[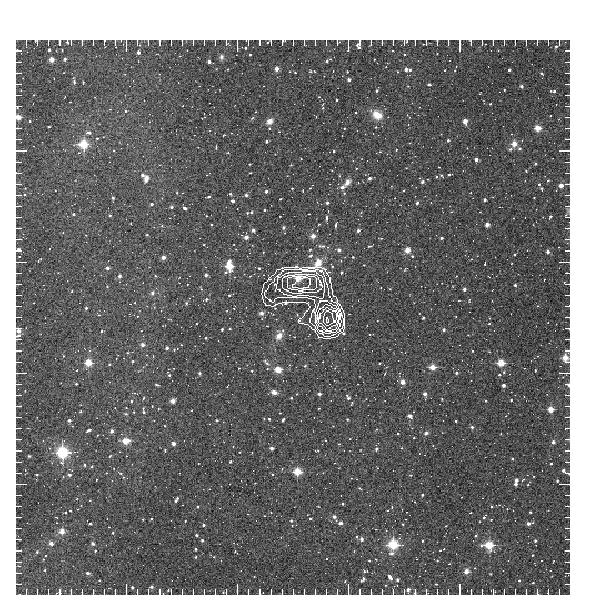]{Abell 2356 R image with X-ray overlay.  Image dimensions are $23.21 \times 23.21$ arcminutes.  The optical image was taken with the 0.9-meter telescope at KPNO; the X-ray image is from Einstein archives.  Contour levels are plotted at 5.0, 6.0, 7.0, 8.0, 9.0, 10.0, and 11.0 counts (Abell 2356 was not included in Forman et al.'s (1981) study). }

\newpage

\begin{figure}[p]
\figurenum{4}
\centerline{\psfig{figure=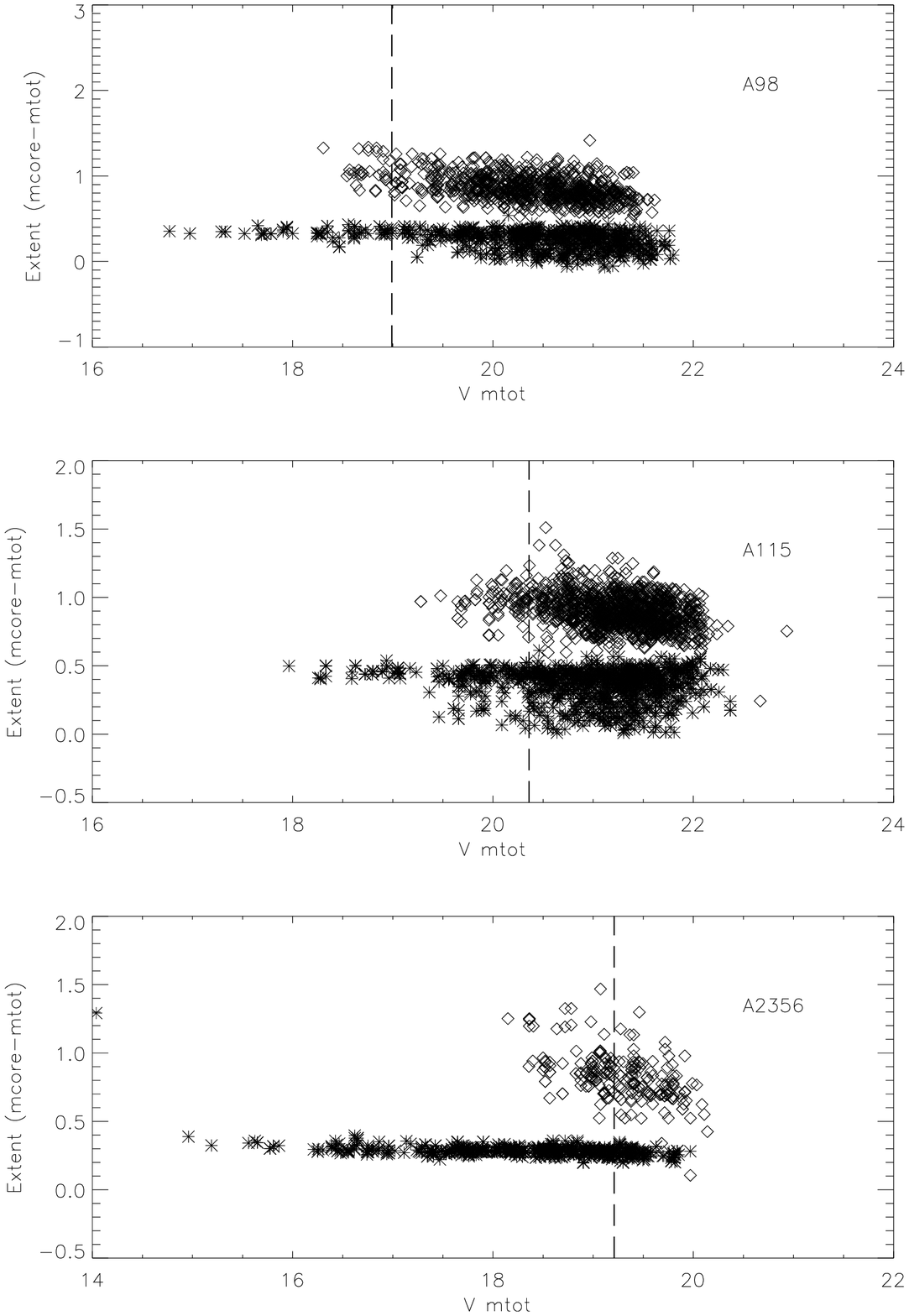,height=6.in}}
\caption{FOCAS star/galaxy classification check.  Here $m_{
core}-m_{tot}$ represents the extent of a detected object.  Galaxies, being extended objects, have $m_{core}-m_{tot}>0$; stars have $m_{core}-m_{tot} \sim 0$.  Because of the large number of data points, the data have been median filtered for clarity.  In this diagram one can see that stars (asterisks) and galaxies (diamonds) are clearly separated to the left of our magnitude cutoff ($M_{v}=-20$).}
\end{figure}

\begin{figure}[p]
\figurenum{5}
\centerline{\psfig{figure=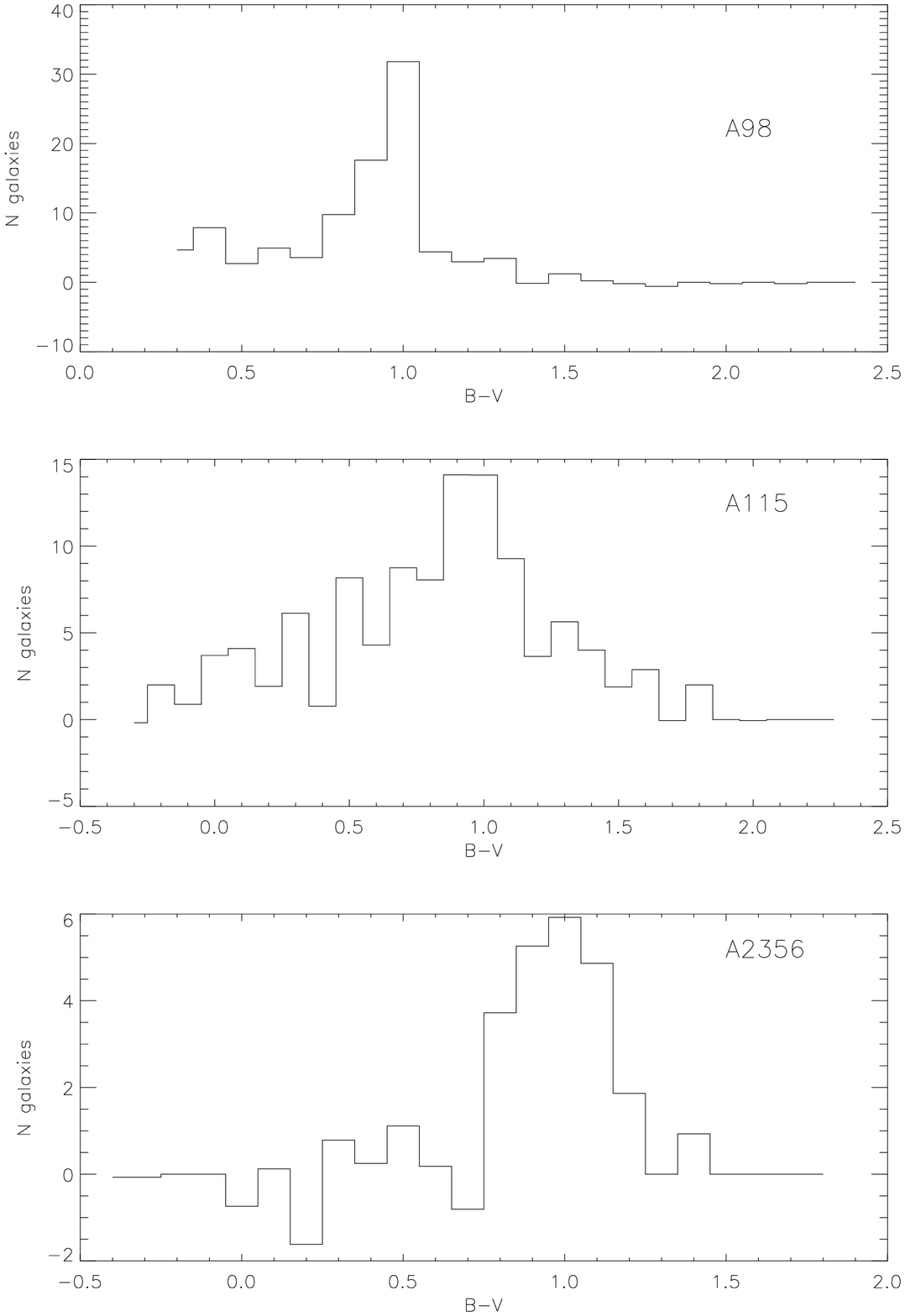,height=6.in}}
\caption{Galaxy color distributions in Abell 98 (top), Abell 115 (middle), and Abell 2356 (bottom).  In each case, the distributions have been background subtracted and color-magnitude corrected.  The histograms have been normalized on the color axis so that majority of the galaxies lie near the E/S0 ridgeline (B--V $\sim$ 1.0).  It is clear that we have well sampled the cluster color distributions.}
\end{figure}

\newpage

\begin{figure}[p]
\figurenum{6}
\centerline{\psfig{figure=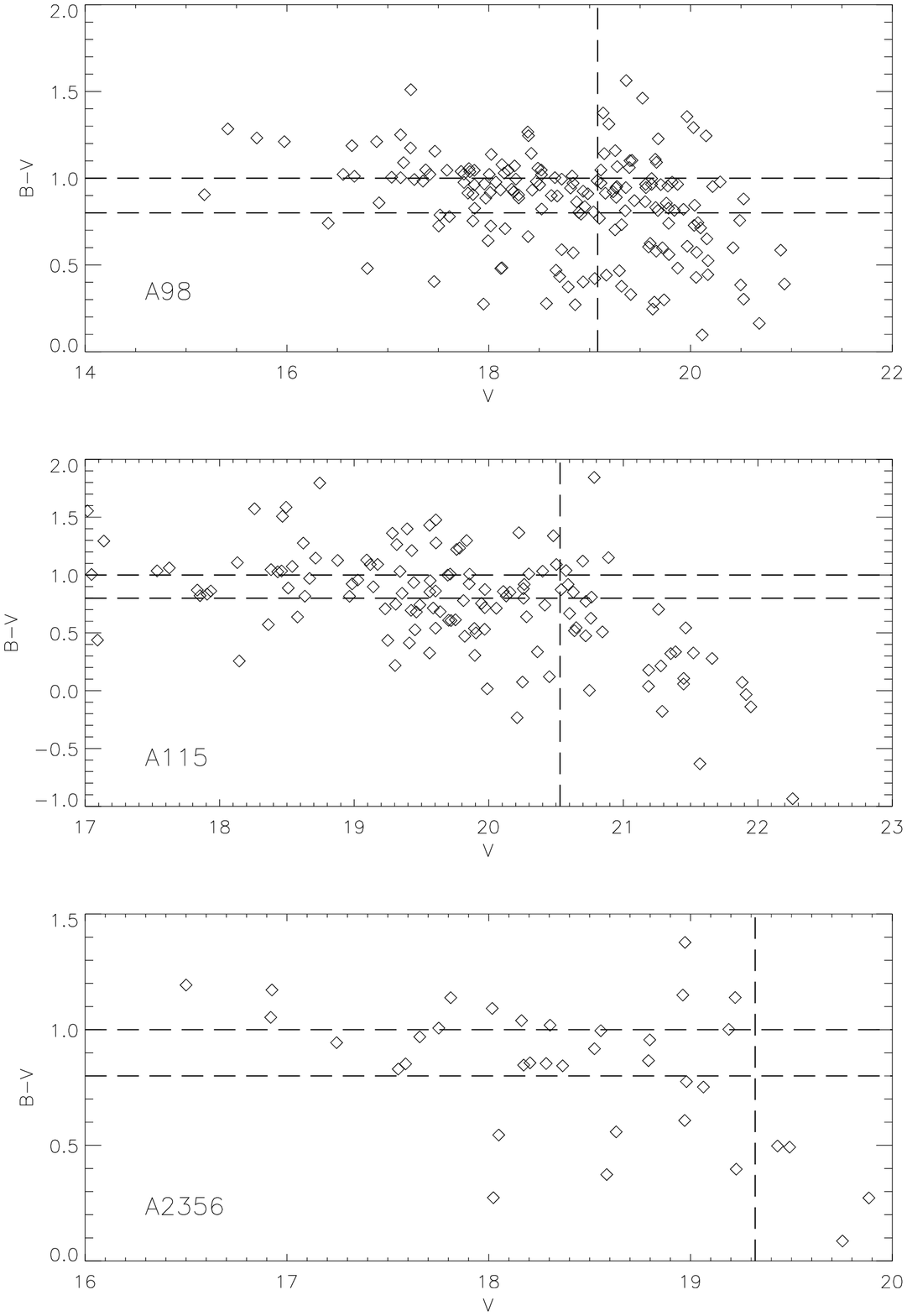,height=6.in}}
\caption{Color-magnitude diagrams for Abell 98 (top), Abell 115 (middle), and Abell 2356 (bottom).  In each case, the distributions have been color-magnitude corrected, but statistical background subtraction has not been done.  Horizontal lines have been drawn at each cluster's E/S0 ridgeline (B--V $\sim$ 1.0) and blue galaxy limit (B--V $\sim$ 0.8).  A vertical line denotes the apparent V magnitude corresponding to $M_{v}=-20$.  Galaxies to the left of this vertical line and below B--V=0.8 are considered blue.}
\end{figure}

\newpage

\begin{figure}[p]
\figurenum{7}
\centerline{\psfig{figure=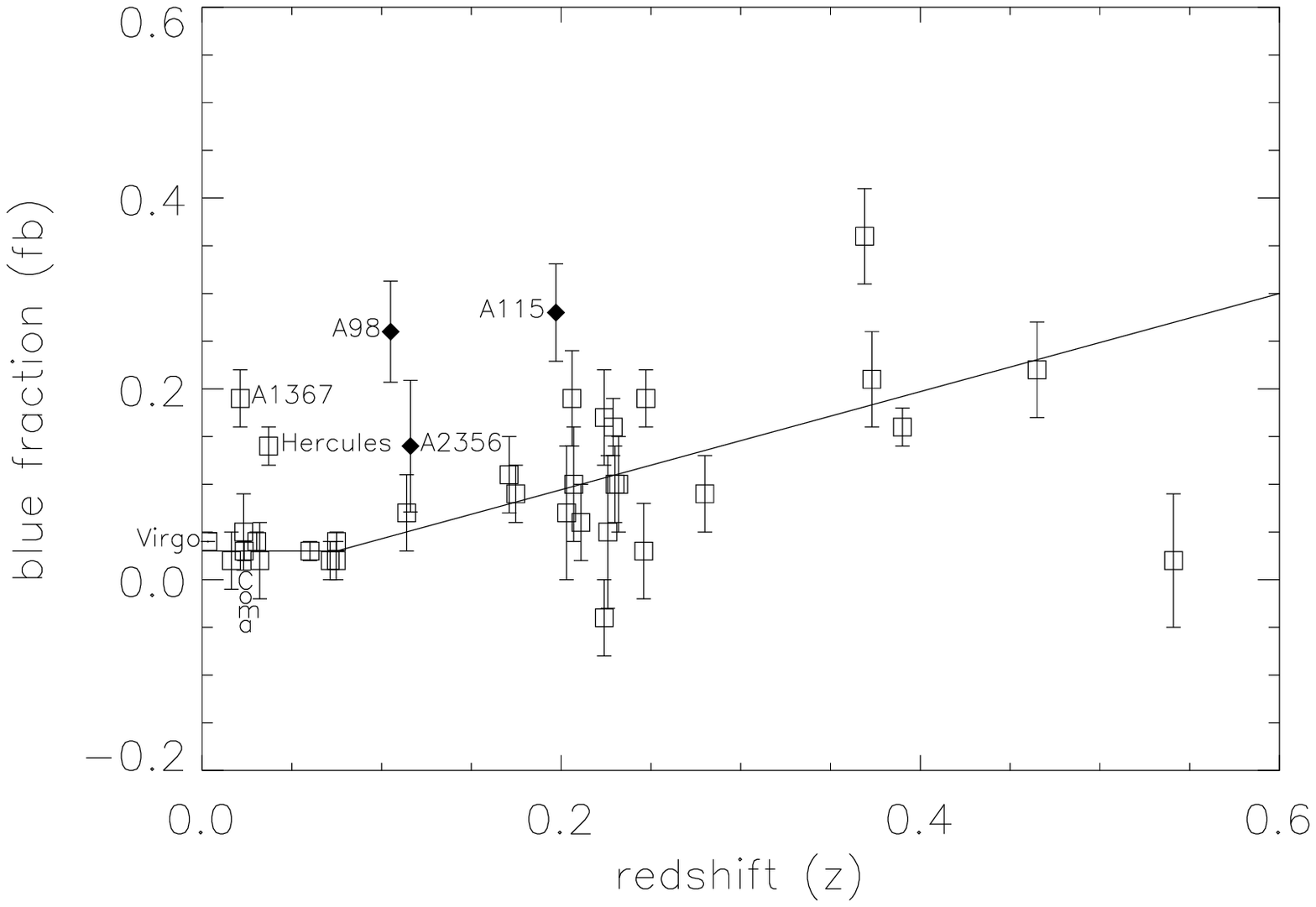,height=4.in}}
\caption{The Abell 98, Abell 115, and Abell 2356 blue fractions we determined (filled diamonds) plotted against Butcher and Oemler (1984) cluster blue fractions (squares).  The BO84 data and ``guiding line" are taken from Table 1 and Figure 3 of BO84.  Fractions of blue galaxies we determined for at least two of our three bimodal clusters are clearly larger than those derived by BO84 for clusters at similar redshifts.}
\end{figure}

\newpage

\begin{figure}[p]
\figurenum{8}
\centerline{\psfig{figure=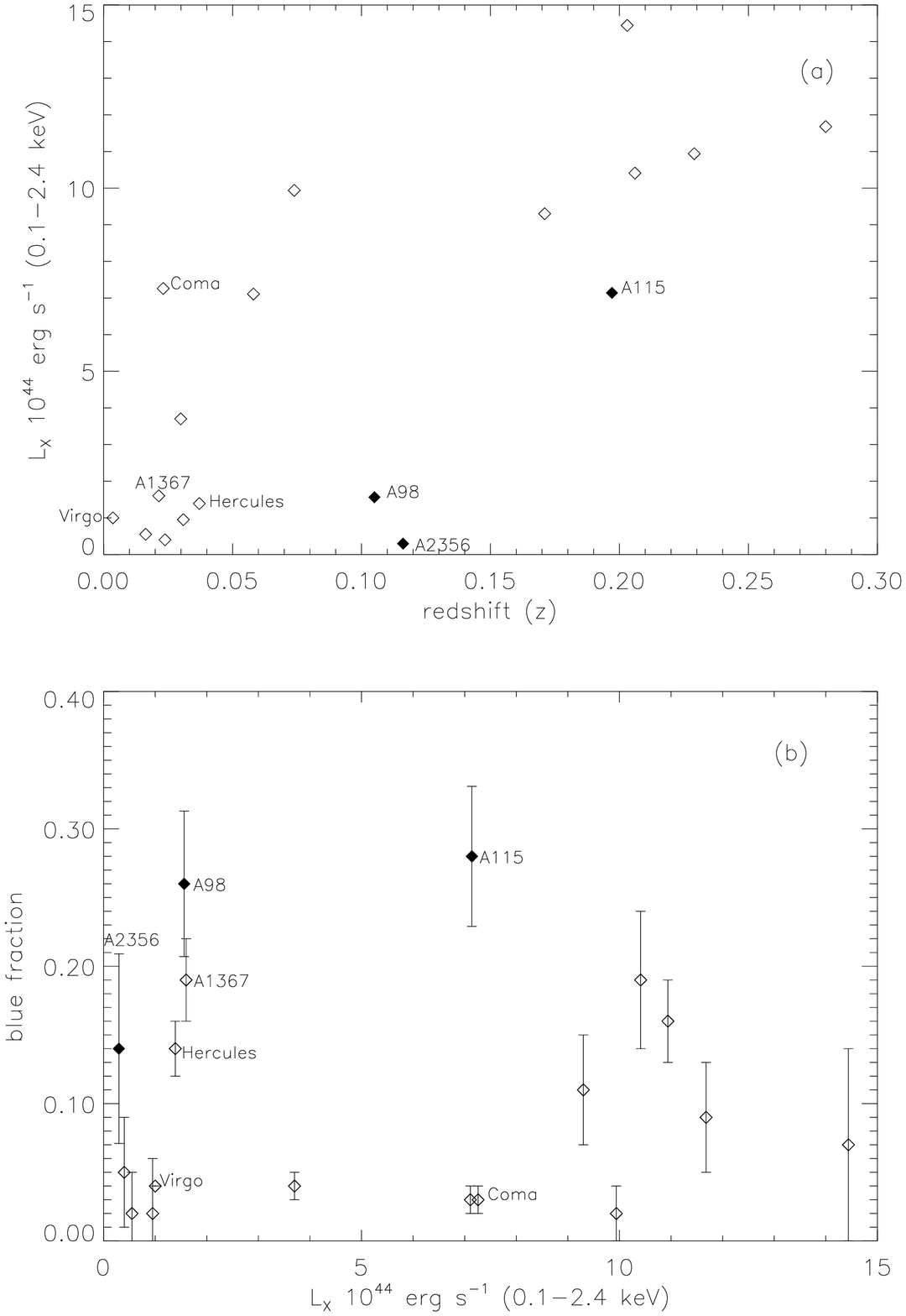,height=6.in}}
\caption{Here we compare the X-ray luminosities of Abell 98, Abell 115, and Abell 2356 (filled diamonds) to those of fifteen Butcher and Oemler (1984) clusters (open diamonds).  BO84 cluster luminosities are from Ebeling et al. (1998).  Notice that cluster X-ray luminosities generally increase with redshift; there is no obvious correlation between $L_{X}$ and cluster blue fraction.}
\end{figure}

\newpage

\begin{deluxetable}{lccc}
\tablewidth{0pt}
\tablecolumns{4}
\tablecaption{Cluster Properties}
\tablehead{\colhead{Cluster} & \colhead{$z$} & \colhead{Abell Richness Class} & \colhead{$L_{X}$\tablenotemark{a} \hspace*{0.1cm} $10^{44}$ erg s$^{-1}$} \\ & & & (0.1-2.4 keV)}
\startdata
Abell 98 & 0.105 & III & 0.900\tablenotemark{b} \hspace*{0.1cm} (N) \\
 & & & 0.662\tablenotemark{b} \hspace*{0.1cm} (S) \\
Abell 115 & 0.1971 & III & 4.480\tablenotemark{c} \hspace*{0.1cm} (N) \\
 & & & 2.660\tablenotemark{c} \hspace*{0.1cm} (S) \\
Abell 2356 & 0.1161 & II & 0.180\tablenotemark{d} \hspace*{0.1cm} (N) \\
 & & & 0.165\tablenotemark{d} \hspace*{0.1cm} (S) \\

\tablenotetext{a}{None of these clusters were observed with ROSAT, so we derived their luminosities in the 0.1-2.4 keV band using XSPEC version 10 (Arnaud 1996) in order to compare to the X-ray luminosities of Butcher-Oemler clusters given in Ebeling et al. (1998; see Figure 8).}
\tablenotetext{b}{Subcluster 1-3 keV luminosity from Forman et al. (1981); our 0.1-2.4 keV $L_{X}$ derivation took into account different $q_{0}$ values used by Forman et al. and Ebeling et al. (1998).  We assumed a temperature of $kT=3$ keV based on the Arnaud \& Evrard (1999) $L$-$T$ relation.}
\tablenotetext{c}{Subcluster 1-3 keV luminosity from Forman et al. (1981); our 0.1-2.4 keV $L_{X}$ derivation was done in the same way as for A98, but here we assumed a temperature of $kT=5.5$ keV.}
\tablenotetext{d}{Subcluster 0.8-3.5 keV luminosity from Ulmer \& Cruddace (1982); our 0.1-2.4 keV $L_{X}$ derivation takes into account different $H_{0}$ values assumed by Ebeling et al. and Ulmer \& Cruddace, as well as the fact that Ulmer \& Cruddace assumed $kT=6.5$ keV for their calculation.  Here we assumed a temperature of 2 keV based on the Arnaud \& Evrard $L$-$T$ relation. }

\enddata
\end{deluxetable}

\newpage

\begin{deluxetable}{lcccc}
\tablewidth{0pt}
\tablenum{2}
\tablecaption{Galactic Extinction Corrections}
\tablehead{\colhead{Cluster} & \colhead{R.A.} & \colhead{Dec.} & \colhead{Gal. Ext. Corr.} & \colhead{Gal. Ext. Corr.}\\
 & & & B mag & V mag }
\startdata
Abell 98 & 00h 46m 13s & 20d 27m 54s & 0.154 & 0.118\\
Abell 115 & 00h 55m 45s & 26d 17m 47s & 0.174 & 0.133\\
Abell 2356 & 21h 35m 33s & 00d 06m 00s & 0.220 & 0.168\\
\enddata
\end{deluxetable}

\begin{deluxetable}{lccc}
\tablewidth{0pt}
\tablenum{3}
\tablecaption{K-corrections}
\tablehead{\colhead{Cluster} & \colhead{z} & \colhead{$k$-corr.} & \colhead{$k$-corr.}\\
 & & B mag & Vmag }
\startdata
Abell 98 & 0.105 & 0.501 & 0.181\\
Abell 115 & 0.1971 & 0.965 & 0.398\\
Abell 2356 & 0.1161 & 0.573 & 0.199\\
\enddata
\end{deluxetable}

\begin{deluxetable}{lccccc}
\tablewidth{0pt}
\tablenum{4}
\tablecaption{Blue Fraction Results} 
\tablehead{\colhead{Cluster} & \colhead{$z$} & \colhead{$N$\tablenotemark{a}} & \colhead{$N_{b}$\tablenotemark{a}} & \colhead{$f_{b}$} & \colhead{Expected BO84 $f_{b}$\tablenotemark{b}} }
\startdata
Abell 98 & 0.105 & 67.7 & 17.3 & $0.26 \pm 0.053$ & $0.046 \pm 0.013$ \\
Abell 98 N & 0.1043 & 6.5 & 3.9 & $0.60 \pm 0.193$ & $0.045 \pm 0.014$ \\
Abell 98 S & 0.1063 & 29.5 & 5.6 & $0.19 \pm 0.073$ & $0.046 \pm 0.014$ \\
Abell 115 & 0.1971 & 76.7 & 21.6 & $0.28 \pm 0.051$ & $0.093 \pm 0.0068$ \\
Abell 2356 & 0.1161 & 25.9 & 3.7 & $0.14 \pm 0.069$ & $0.051 \pm 0.013$ \\
 & & & & & \\
Abell 1367 & 0.0213 & ----- & ----- & $0.16 \pm 0.03\tablenotemark{c}$ & $0.030 \pm 0.0044$ \\
Hercules & 0.0371 & ----- & ----- & $0.14 \pm 0.02\tablenotemark{c}$ & $0.030 \pm 0.0044$ \\
\tablenotetext{a}{Background-subtracted values.}
\tablenotetext{b}{Blue fractions predicted from the BO84 line fit (see Figure 7).}

\tablenotetext{c}{Data taken from BO84, Table 1; errors for expected blue fractions calculated by the authors of this paper.}

\enddata
\end{deluxetable}

\end{document}